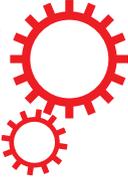

OPEN

# ceRNA crosstalk stabilizes protein expression and affects the correlation pattern of interacting proteins

Araks Martirosyan[1], Andrea De Martino[2,3,4,*], Andrea Pagnani[3,5,*] & Enzo Marinari[1,6,*]

Gene expression is a noisy process and several mechanisms, both transcriptional and post-transcriptional, can stabilize protein levels in cells. Much work has focused on the role of miRNAs, showing in particular that miRNA-mediated regulation can buffer expression noise for lowly expressed genes. Here, using *in silico* simulations and mathematical modeling, we demonstrate that miRNAs can exert a much broader influence on protein levels by orchestrating competition-induced crosstalk between mRNAs. Most notably, we find that miRNA-mediated cross-talk (i) can stabilize protein levels across the full range of gene expression rates, and (ii) modifies the correlation pattern of co-regulated interacting proteins, changing the sign of correlations from negative to positive. The latter feature may constitute a potentially robust signature of the existence of RNA crosstalk induced by endogenous competition for miRNAs in standard cellular conditions.

The control of gene expression noise is a challenge faced, at some degree, by cells in all organisms, as each post-transcriptional step of the genetic regulatory chain, including translation alone[1], can potentially amplify transcription noise. Biological functionality however often requires finely-tuned protein levels. Cells therefore employ a variety of strategies to ensure that protein noise is buffered[2–9]. Regulatory RNAs, and microRNAs (miRNAs) in particular, are thought to play a major role in this respect[10,11].

miRNAs comprise a large number of short, endogenously expressed non-coding RNA species that are significantly conserved among invertebrates and vertebrates and whose expression is strongly tissue-specific[12]. They act primarily as negative controllers of gene expression, by silencing translation and/or catalyzing mRNA destabilization after sequence-specific binding to their targets. They can however also bind non-coding RNA species like pseudogenes and long non-coding RNAs (lncRNAs)[13,14]. In some cases, 'sponging' of miRNAs by lncRNAs has been found to contribute significantly to the adjustment of miRNA levels in the cell[15]. Overall, miRNAs appear today as key regulators of a very broad class of RNA molecules.

Recent work, both experimental and *in silico*, has started to uncover the complex and highly heterogeneous network of miRNA-RNA interactions, showing that hundreds of genes may be directly repressed by individual miRNAs, albeit modestly in some cases[16–21]. The search for a possible functional rationale for the breadth of miRNA-RNA couplings has increasingly focused on containment of gene expression noise[20–25]. Several studies have investigated the effects of miRNA-mediated regulation on RNA and protein levels[24–26]. Most recently, integrated theoretical and experimental work[24] has shown that miRNA control stabilizes the levels of lowly expressed genes while having the opposite effect on highly expressed ones. This scenario, which mirrors that obtained for miRNA-mediated regulation of transcript levels[27,28], suggests that other mechanisms may confer robustness at high expression levels or, more generally, across the entire expression range. Here we argue that the recently hypothesized 'ceRNA effect'[29] might constitute such a mechanism.

[1]Dipartimento di Fisica, Sapienza Università di Roma, Rome, Italy. [2]Soft & Living Matter Lab, Istituto di Nanotecnologia (NANOTEC-CNR), Rome, Italy. [3]Human Genetics Foundation, Turin, Italy. [4]Center for Life Nano Science@Sapienza, Istituto Italiano di Tecnologia, Rome, Italy. [5]Dipartimento di Scienza Applicata e Tecnologia, Politecnico di Torino, Turin, Italy. [6]INFN, Sezione di Roma 1, Rome, Italy. *These authors contributed equally to this work. Correspondence and requests for materials should be addressed to A.M. (email: araks.martirosyan@uniroma1.it) or A.D.M. (email: andrea.demartino@roma1.infn.it)

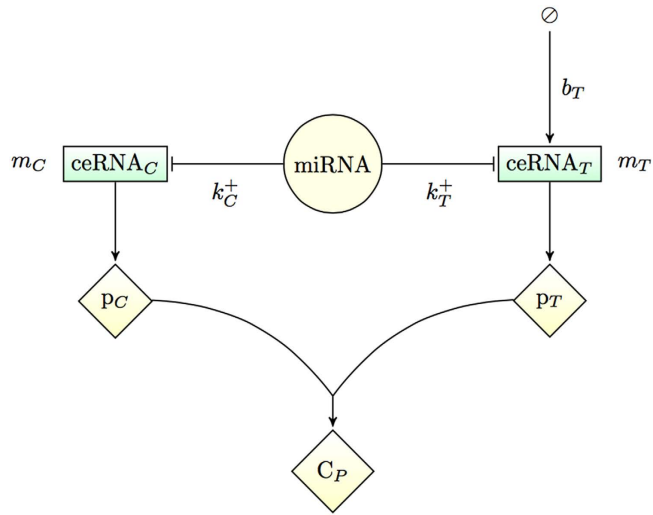

**Figure 1. Model schematics and main parameters.** A single miRNA species negatively controls the expression of the ceRNA species ceRNA$_T$ ('target', level $m_T$) and ceRNA$_C$ ('competitor', level $m_C$), to which it can bind with rates $k_T^+$ and $k_C^+$, respectively. The functional products of the ceRNAs, proteins $p_T$ and $p_C$, can eventually interact to form a complex $C_P$. The key control parameter is the target's transcription rate $b_T$. See Fig. 7 for a detailed scheme that includes all processes.

The ceRNA effect (whereby ceRNA stands for 'competing endogenous RNAs') consists, in essence, of a positive effective interaction that can arise between transcripts that are targeted by the same miRNA species due to competition[29–32]. The ability of the ceRNA mechanism to mediate RNA cross-talk has been investigated in detail *in silico*[27,28,33], while experimental validations have been obtained by over-expressing miRNAs or targets[16] and in specific conditions related to disease or differentiation[15,34,35]. Its relevance in standard physiological conditions is less clear[16,17]. On one hand, effective ceRNA cross-talk appears to depend strongly on the kinetics of miRNA-target interactions, on gene silencing mechanisms and on the relative abundance of regulators and targets, and therefore may require considerable fine tuning[27,28,36,37]. On the other, miRNAs can achieve optimal or nearly optimal regulation of competing transcripts in a broad range of parameter values by exploiting heterogeneities in e.g. binding kinetics or target degradation pathways[36]. A natural question is whether proteins, i.e. the functional products of coding transcripts, may also benefit from miRNA-mediated control when the respective mRNAs are competing to bind miRNAs. Interestingly, studies of protein-protein interaction (PPI) networks have revealed that interacting proteins tend to be regulated by miRNA clusters, i.e. by co-expressed miRNAs[37–41]. More specifically, it was found that (a) individual or co-expressed miRNAs frequently target the mRNAs of several components of protein complexes[38], (b) direct miRNA targets and their partners show significant modularity at the level of the corresponding proteins in the human PPI network[39], and (c) the products of transcripts targeted by the same miRNAs are more connected in the human PPI network than expected by chance[40]. These observations suggest that miRNAs may be implicated in the fine control of interacting proteins, making the ceRNA effect a rather natural mechanism through which such control could be exerted.

Here we quantify the impact of ceRNA competition on protein fluctuations and on protein complex formation by mathematical modeling and *in silico* experiments. Within a minimal stochastic description that includes both transcriptional and post-transcriptional control, we show that:

(a) ceRNA cross-talk can stabilize the level of highly expressed proteins (with respect to the case in which no competition takes place);
(b) the ceRNA effect alters the correlation pattern of co-regulated interacting proteins, particularly by turning its sign from negative to positive;
(c) miRNA recycling enhances the suppression of protein expression noise through the ceRNA effect across the entire range of expression levels.

These results have significant implications. First, they suggest that ceRNA cross-talk may be crucial for the fine tuning of protein levels, thereby pointing to a further explanation for the abundance of lncRNAs and pseudogenes (i.e. of miRNA 'sponges') in the human transcriptome. Secondly, they indicate that a positive correlation between co-regulated subunits of a protein complex may provide, for a limited but significant set of cases, the simple and direct proof of active ceRNA cross-talk *in standard physiological conditions* that has been so far lacking.

### Results

**Model description and basic properties.** As a basis, we consider the model of a ceRNA network studied previously in[27,28,36], with the addition of protein synthesis and a protein complex formation step (see Fig. 1). In short, a miRNA species negatively regulates the expression of two ceRNAs, whose levels are denoted respectively as $m_T$ (for 'target') and $m_C$ (for 'competitor'). Both serve as substrates for protein synthesis and the respective products ($p_T$ and $p_C$) can interact to form a complex ($C_P$). By turning specific parameters on or off the above basic model allows to tackle different situations. A detailed description of the model, including the various parameters, is given in the Methods. Its dynamics, formalized in terms of stochastic differential equations, can be simulated using the Gillespie algorithm[42], whereas analytical estimates for quantities like correlation functions can be obtained via the Linear Noise Approximation[43] (LNA); see Methods for details.

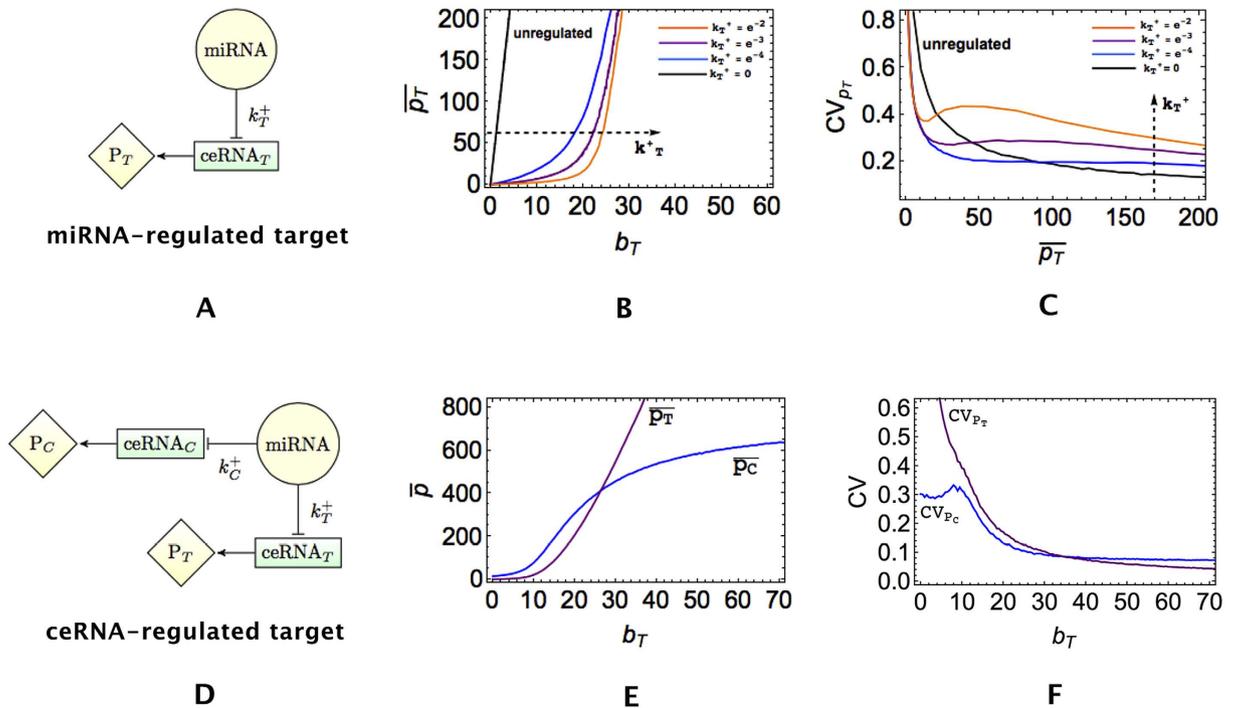

**Figure 2. Dependence of mean protein expression levels and relative fluctuations (CV) on the transcription rate $b_T$ of the target.** Panels (B) and (C) describe the case of a simple miRNA-regulated target, shown in panel (A). In (B), $k_T^+$ increases in the direction of the arrow (specifically, $k_T^+ = e^{-2}, e^{-3}, e^{-4}, 0$ for orange, purple, blue, black curves respectively). Panels (E) and (F) describe the case of a target regulated through ceRNA competition, depicted in (D), for $k_T^+ = e^{-2}$ and $k_C^+ = e^{-3}$. Note that no PPI is considered in this case.

The basic features of miRNA-mediated regulation that emerge from this model at stationarity with and without ceRNA competition (and in absence of PPI) are summarized in Fig. 2. In absence of ceRNA competition (Fig. 2A–C), upon varying the synthesis rate of the target ($b_T$) while keeping all other parameters fixed, the expression of the target's functional product $p_T$ undergoes a crossover from a repressed regime with low copy numbers to a free (unrepressed) regime in which its level increases roughly linearly with $b_T$. The crossover gets sharper as the miRNA-target interaction strength $k_T^+$ increases. The ability of miRNAs to generate threshold-linear expression profiles via molecular titration was pointed out already in several studies[27,28,44,45]. In the crossover region, $p_T$ displays strong sensitivity to small changes in $b_T$, as testified by peaks in the coefficient of variation (CV) that get more marked and shift to lower values of protein concentration as $k_T^+$ increases, see Fig. 2C. In this regime, called 'susceptible' in Figliuzzi *et al.*[28], miRNA and mRNA levels are nearly equimolar[27,28,34]. Contrasting this behaviour with the standard Poissonian CV obtained for an unregulated protein ($k_T^+ = 0$, black line in Fig. 2B and C) one sees that miRNA control generically buffers expression noise for lowly expressed genes while it amplifies noise for highly expressed genes, in agreement with recent experimental work quantifying protein expression noise in miRNA-regulated genes[24]. When a competitor is present (Fig. 2D–F), one observes that, upon modulating $b_T$, the level of $p_C$ starts changing when $b_T$ is around the crossover region, reflecting the effective positive coupling between ceRNAs $m_T$ and $m_C$ known as the 'ceRNA effect'. The emergence and major features of the ceRNA effect at the level of transcripts have been characterized in refs 27,28,33,36,46,47.

We shall now analyze in more detail the influence of miRNA sponging and ceRNA competition on protein expression noise and on the PPI. Following[36,48,49], the effectiveness of the regulatory channel linking an input variable, $b_T$ in this case, to an output variable $O$ (e.g., the target protein level $p_T$ or the level of the protein complex $C_P$) will be characterized by its capacity $I_{max}$, defined as the maximum mutual information between $b_T$ and $O$ that can be achieved by changing the input distribution $p(b_T)$ while keeping the conditional distribution $p(O|b_T)$ (the 'channel', which stochastically returns a value of $O$ upon presenting input $b_T$) fixed:

$$I_{max} = \max_{p(b_T)} I(b_T, O), \quad I(b_T, O) = \int_{b_T^{min}}^{b_T^{max}} db_T p(b_T) \int_{O^{min}}^{O^{max}} dO\, p(O|b_T) \log_2 \frac{p(O|b_T)}{p(O)} \quad (1)$$

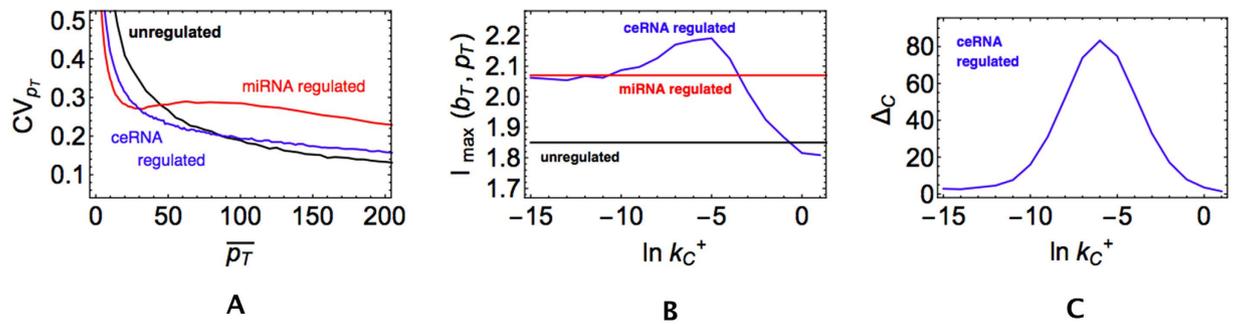

**Figure 3. ceRNA competition can stabilize highly expressed proteins. (A)** Coefficient of variation of $p_T$ as a function of the mean protein level for a post-transcriptionally unregulated protein (black line, $k_C^+ = 0$ and $k_T^+ = 0$), a miRNA-regulated protein (red line, $k_C^+ = 0$ and $k_T^+ = e^{-3}$) and a ceRNA-regulated protein (blue line, $k_C^+ = e^{-5}$ and $k_T^+ = e^{-3}$). **(B)** Capacity of the target's expression channel as a function of the miRNA-competitor interaction strength. Color code same as in panel A. **(C)** Derepression size $\Delta_C$ of the competitor as a function of the miRNA-competitor interaction strength in the case of ceRNA regulation (same parameters as panel B).

where $p(O) = \int_{b_T^{\min}}^{b_T^{\max}} db_T p(O|b_T) p(b_T)$ is the output distribution. We will work under the assumption that $p(O|b_T)$ is Gaussian and its variance is small for each $b_T$ ('Small Noise Approximation'), in which case the above problem has been shown to have a simple analytical solution[49,50] (see the protocol for computing capacities in Methods). Note that the input variable is constrained to vary between fixed bounds $b_T^{\min}$ and $b_T^{\max}$ and that, correspondingly, the output varies between $O^{\min}$ and $O^{\max}$.

**ceRNA cross-talk enhances the stability of highly expressed proteins.** Recent experimental work[24] has revealed that, while miRNA regulation suppresses protein noise for lowly expressed genes, at high expression levels noise is generically larger. This effect can be surprisingly reversed through ceRNA competition. In particular, by titrating miRNA molecules away from their target, a competitor can enhance the target's stability. This is shown in Fig. 3A, where the CVs of proteins that are post-transcriptionally unregulated, regulated by a miRNA and regulated by the ceRNA effect are compared. One sees that the relative fluctuations of the target at high expression levels can be substantially reduced by the onset of ceRNA competition with respect to the miRNA-regulated case without significantly affecting the low expression regime. In addition, ceRNA regulation generates a fluctuation scenario similar to the Poissonian one that characterizes an unregulated protein, showing that competition buffers noise essentially by de-repressing the target.

In Fig. 3B we show how the capacity of the protein expression channel changes with the strength $k_C^+$ of the interaction between the competitor, ceRNA$_C$, and the miRNA. For small $k_C^+$, $I_{\max}(p_T, b_T)$ expectedly tends to the value obtained for a miRNA-regulated protein as the effect of the competitor gets weaker and weaker. For large $k_C^+$, on the other hand, the competitor tends to sponge all miRNAs away from the target, therefore leading to a capacity close to that of a simple transcriptional control unit. For intermediate values of $k_C^+$, instead, $I_{\max}$ peaks, signalling that the expression of $p_T$ can be tuned more efficiently than by transcriptional or miRNA-mediated regulation alone. That the ceRNA effect is at origin of this behaviour can be checked by measuring the derepression size of ceRNA$_C$, $\Delta_C$, defined as the difference between the largest and smallest steady state values of $m_C$ that are obtained by changing the input variable $b_T$ between its smallest and largest allowed values $b_T^{\max}$ and $b_T^{\min}$:

$$\Delta_C = m_C(b_T^{\max}) - m_C(b_T^{\min}). \tag{2}$$

Based on the ceRNA effect features shown in Fig. 2E, a large derepression size indicates that the ceRNA effect is active. Clearly, see Fig. 3C, $\Delta_C$ is markedly larger for intermediate values of $k_C^+$, thereby pointing to the onset of ceRNA crosstalk.

**miRNA recycling increases the potential of miRNA-mediated gene regulatory circuits to suppress gene expression noise.** miRNA-ceRNA complexes can be processed in different ways, including via unbinding and complex degradation ('stoichiometric decay')[12,51,52]. However, if miRNAs have a perfect complementarity with the target, catalytic cleavage of the latter can be induced[53–57], after which miRNAs are likely to be available again for interaction with a new target ('miRNA recycling'). This catalytic channel of complex processing may effectively increase the population of free miRNA regulators with respect to targets, thereby enabling a more subtle control of gene expression[36]. Figure 4 elucidates its role in controlling protein levels. By increasing the efficiency of mRNA cleavage at the target node (i.e. for faster rates of catalytic processing $\kappa_T$, see Methods), protein expression noise improves significantly with respect to the case of slow catalytic processing. Notice that the same effect is obtained in a simpler miRNA-regulated element. In specific, while for a ceRNA regulated target the stability of the protein levels improves by about 20%, for a miRNA-regulated target the improvement may be more relevant – up to about 40%.

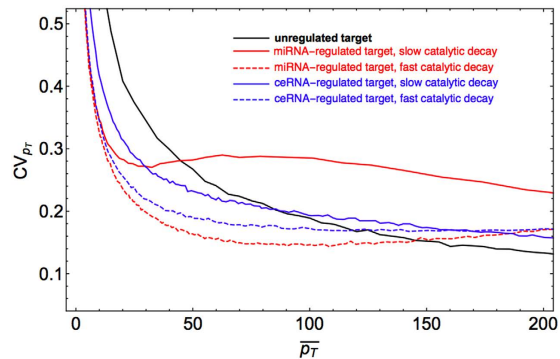

**Figure 4. Noise reduction in miRNA-mediated circuits due to the effective miRNA-recycling at the target node.** Fast and slow (high and low $\kappa_T$, see Methods) miRNA recycling scenarios at the target node are shown, respectively, by dashed and solid lines for a miRNA-regulated protein (red, $k_C^+ = 0$ and $k_T^+ = e^{-3}$) and a ceRNA-regulated protein (blue, $k_C^+ = e^{-5}$ and $k_T^+ = e^{-3}$). The black curve describes the case of an unregulated target ($k_C^+ = 0$ and $k_T^+ = 0$).

### The ceRNA effect alters the correlation pattern of co-regulated interacting proteins.

Competition to bind regulatory miRNAs can lead to a positive correlation among co-regulated transcripts, who can respond both to fluctuations in miRNA levels and, at fixed miRNA level, to changes in ceRNA levels[27–29,36]. This is shown in Fig. 5A, where one also sees that the corresponding (non-interacting) proteins have a similar correlation pattern, as measured by the Pearson coefficient between the levels of free molecules. Proteins that form a complex are instead negatively correlated by the PPI in absence of co-regulation at the level of transcripts (in which case transcripts are uncorrelated), as shown in Fig. 5B. The negative correlation is caused by the protein complex binding kinetics alone. Interestingly, the ceRNA effect can reverse this scenario, i.e. in presence of upstream miRNA competition interacting proteins can become positively correlated (Fig. 5C). In particular, the Pearson coefficient $\rho(p_T, p_C)$ is largest close to the regime where the ceRNA effect is strongest and $\rho(m_T, m_C)$ peaks, while it becomes negative when the ceRNA-ceRNA correlation weakens.

Several proteins forming binary complexes and known to share miRNA regulators display a positive correlation. For instance, both subunits of the CCND1:CDK4 complex are regulated by miR-545 in human[58]. Experiments in laryngeal squamous cells have revealed that, under CCND1 over-expression, the expression level of CDK4 increases[59]. Likewise, ITGA6 down-regulation results in ITGB4 decrease in cells where the ITGA6:ITGB4 complex is present[60]. Both ITGA6 and ITGB4 are known to share common miRNA regulators in humans[38]. Based on the above results, a positive correlation between subunits of a protein complex may therefore be explained in terms of miRNA-mediated cross-talk at the level of the respective transcripts.

Incidentally, we note that, as shown in Fig. 5C, the onset of a positive correlation between target ($p_T$) and competitor ($p_C$) proteins occurs only in a narrow window of parameters around the quasi-equimolar[27] or susceptible[28] regime. It is somewhat striking that the experimental studies of regulated binary complexes discussed here[58–60] report a positive correlation between subunits. This suggests that, at least for these systems, kinetic parameters could be globally tuned by evolution so as to fit this narrow window. A potential added advantage of such a scenario lies, as we shall now see, in the buffering of protein complex noise that can accompany it.

### Co-regulation by a miRNA can effectively control the level of a binary protein complex. 

If protein fluctuations get correlated due to the ceRNA effect, it might be possible to exploit the same mechanism to fine tune the level of the protein complex $C_p$. Figure 6 shows that the capacity of the protein complex synthesis channel is weakly modulated by the miRNA-ceRNA binding rate that quantifies the strength of miRNA regulation on the competitor (see Fig. 3B). On the other hand, optimal regulation is achieved at intermediate values of $k_T^+$, where $I_{\max}$ is slightly above the value obtained for low values of $k_T^+$ and $k_C^+$ (corresponding to the case of unregulated transcript). Interestingly, though, the channel capacity seems to become generically larger as $k_C^+$ grows, suggesting that, under stronger competition, a more efficient tuning of the complex level may be achieved in a broader range of values of $k_T^+$.

### Discussion

A wide variety of biological functions has been assigned over time to non-coding RNA molecules. Most interestingly, perhaps, they can regulate gene expression at the post-transcriptional stage. Eukaryotic miRNAs are post-transcriptional micromanagers of gene expression[61] that exert regulatory roles in situations as different as neuronal regulation[62,63], brain morphogenesis[64], muscle cell differentiation[65], stem cell division[66], glucose and lipid metabolism[67] as well as in many disease states. The identification of viral miRNAs furthermore suggests that viruses may use them to interfere with the host's gene expression[68]. Therapeutics targeting miRNA levels therefore appear to be particularly promising tools. Their viability however has to be evaluated against the broad microenvironment in which miRNAs operate. It is now clear that miRNAs and their targets are linked in a complex transcriptome-scale interaction network and that the effective strength of miRNA-induced repression depends tightly on miRNA-RNA binding and unbinding free energies (that are predicted to show a wide spectrum[69]) and on molecular levels. In such a context, understanding how perturbations probing one node may propagate to other nodes, thereby affecting gene expression as a whole, is far from trivial. Moreover, pseudogenes and long non-coding RNAs can compete with coding transcripts to bind miRNAs[14], and the effects induced by competition can be especially hard to quantify. Indeed, the establishment of an effective positive interaction between different targets of the same miRNAs (whose effectiveness is supported by several perturbation-based experimental studies) may potentially contribute significantly to both the overall gene expression profile and, dynamically, to the re-shaping of the proteome in response to perturbations. While many of its features are well understood[27,28,36,47,70], its significance *in vivo* is debated[16–18]. On the other hand, miRNAs have been shown to be able to confer precision to expression levels either by themselves or in combination with spefic motifs[25,26]. Recent experiments have also shown that, in cells with low expression of the reporter carrying a binding site for the miRNA protein, protein noise was reduced compared to a control, while fluctuations were increased for highly expressed reporters[24].

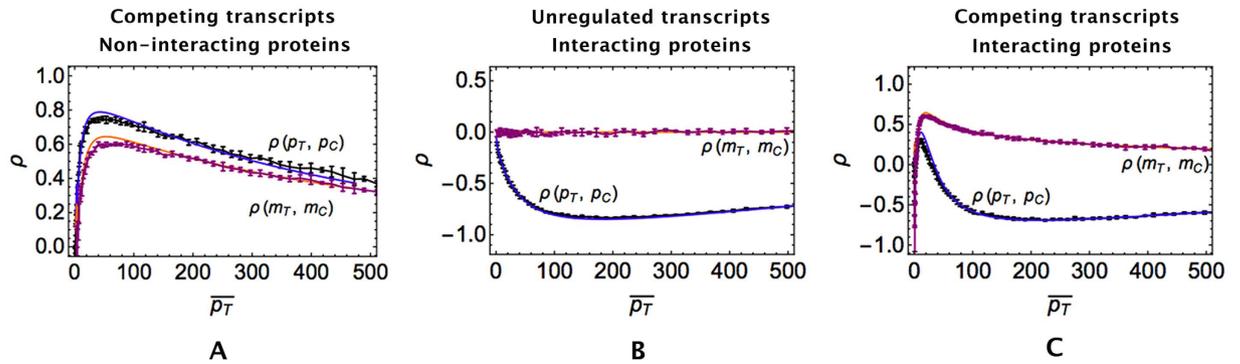

**Figure 5. Correlation patterns for interacting proteins measured by the Pearson coefficient $\rho$ as a function of the mean target level.** (**A**) Case of non-interacting proteins translated from competing transcripts ($k_T^+ = e^{-2}$ and $k_C^+ = e^{-3}$). (**B**) Case of interacting proteins translated from post-transcriptionally unregulated transcripts ($k_T^+ = k_C^+ = 0$). (**C**) Case of interacting proteins translated from competing transcripts ($k_T^+ = e^{-2}$ and $k_C^+ = e^{-3}$). Lines (blue and orange) correspond to analytical results obtained by the Linear Noise Approximation (see Methods), markers (black and purple) to results from stochastic simulations by the Gillespie algorithm.

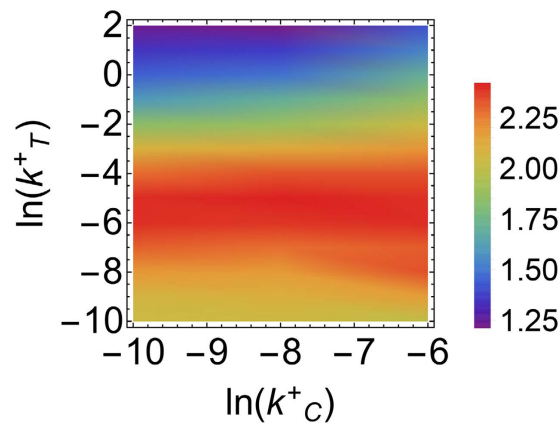

**Figure 6. Capacity of the protein complex synthesis channel, $I_{max}(C_p, b_T)$, as a function of the miRNA-ceRNA binding strengths of the target ($k_T^+$) and the competitor ($k_C^+$).** To allow for comparisons, simulations were performed at fixed output variation range $\Delta_C = C_P^{max} - C_P^{min} = 200$ (with $C_P^{min} = 0$) and variable $b_T$.

Our study establishes a link between different miRNA-mediated functions by showing that buffering of protein expression noise can be enhanced by competition. In specific, competition by itself reduces noise on highly expressed genes, while catalytic processing of the miRNA-ceRNA complex provides a further degree of freedom through which noise can be controlled. As was found to be the case for other emergent properties of miRNA-based regulation[28,36], these effects are enhanced by kinetic heterogeneities, providing further insight into the possible functions served by the evolutionarily-selected, broad distribution of rates found in miRNA-ceRNA networks. In addition, for interacting proteins whose mRNAs are regulated by the same miRNA (as frequently found in actual regulatory networks[38]), our study unveils several new features. First, we argued that, due to the strong ceRNA effect, a positive correlation may be established between the two sub-units of a complex, reversing the negative sign that is induced by complex formation in absence of co-regulation by a common modulator. This is a direct consequence of ceRNA cross-talk and may provide a first, important signature of ceRNA cross-talk *in vivo*.

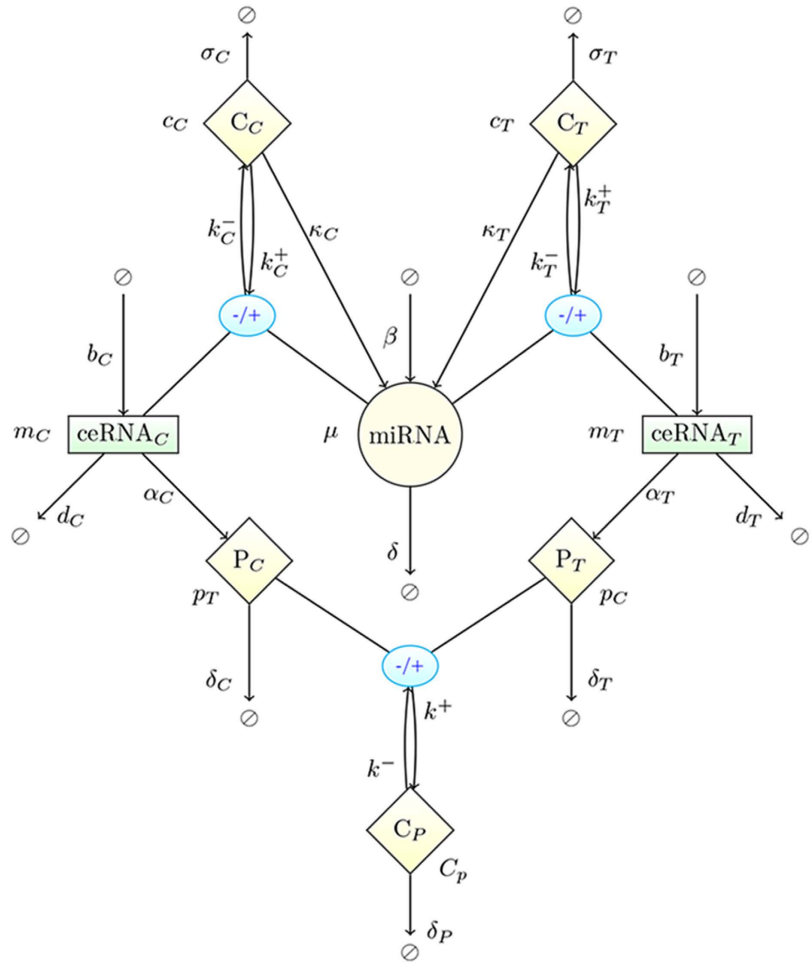

**Figure 7. Schematic representation of the model.** Two proteins ($p_T$ and $p_C$) translated from 2 distinct ceRNAs ($m_T$ and $m_T$) that are regulated by the same miRNA ($\mu$). Proteins $p_T$ and $p_C$ associate and dissociate to a protein complex $C_p$ with a rate $k^+$ and $k^-$ respectively, miRNA binds (unbinds) to the ceRNAs $m_T$ and $m_C$ with the rates $k_T^+(k_T^-)$ and $k_C^+(k_C^-)$ respectively forming ceRNA:miRNA complexes $c_T$ and $c_C$. Species $m_T$, $m_C$, $\mu$, $p_T$, $p_C$ are synthesized (degraded) with the rates $b_T$, $b_C$, $\beta$, $\alpha_T$, $\alpha_C$ ($d_T$, $d_C$, $\delta$, $\delta_T$, $\delta_C$) correspondingly. Protein complex $C_p$ undergoes spontaneous degradation with a rate $\delta_p$. Finally, $c_T$ and $c_C$ decay catalytically (ceRNA cleavage and miRNA recycling) with the rates $\kappa_T$ and $\kappa_C$ respectively. Figure-specific values of the kinetic parameters are reported in Table 1.

Finally, we have shown that protein complexes may be synthesized with higher precision if its components undergo miRNA-mediated post-transcriptional regulation, although the effect can be modest.

The influence of miRNAs on protein complexes has been experimentally investigated in the context of diseases. For instance, analyses of miRNA-mediated dysregulation of functionally related proteins during prostate cancer progression had identified miRNA-1 and miRNA-16 as master regulators of prostate cancer, since they regulate hubs of the underlying PPI network – the SMAD4 and HDAC proteins[71]. Other studies have revealed the regulatory role of miR141–200c in the epithelial-to-mesenchymal transition due to the orchestrated regulation of the CtBp/Zeb complex[38]. Likewise, miR-200 has been identified as a powerful marker of the epithelial phenotype of cancer cells[72], while miR141–200c targets $\beta$-catenin, the downstream effector of the Wnt proliferation pathway[73], and miR-545 targets complex-forming CCND1 and CDK4 with potentially suppressive effects on the proliferation of lung cancer cells[58]. The present work established an *in silico* framework through which such cases could be quantitatively analyzed by probing the roles of the various parameters that regulate miRNA regulation, competition, cross-talk and PPI. Experimental validation in perturbation-based experiments would provide further understanding on the global effectiveness of miRNAs as controllers of cellular protein profiling.

## Methods

**Stochastic model.** The dynamics of the model, presented in detail in Fig. 7, can be described by the stochastic equations

| Parameter | Description | All Figures | | | |
|---|---|---|---|---|---|
| $b_T$ [molecule min$^{-1}$] | Target synthesis rate | [0, 60] | | | |
| $b_C$ [molecule min$^{-1}$] | Competitor synthesis rate | 15 | | | |
| $\beta$ [molecule min$^{-1}$] | miRNA synthesis rate | 25 | | | |
| $k_T^+$ [molecule$^{-1}$ min$^{-1}$] | Target:miRNA complex association rate | see captions | | | |
| $k_T^-$ [min$^{-1}$] | Target:miRNA complex dissociation rate | 0.001 | | | |
| $k_C^-$ [min$^{-1}$] | Competitor:miRNA complex dissociation rate | 0.001 | | | |
| $\kappa_C$ [min$^{-1}$] | Competitor:miRNA complex catalytic decay rate | 0.001 | | | |
| $d_T$ [min$^{-1}$] | Target decay rate | 0.1 | | | |
| $d_C$ [min$^{-1}$] | Competitor decay rate | 0.1 | | | |
| $\sigma_T$ [min$^{-1}$] | Target:miRNA complex decay rate | 1 | | | |
| $\sigma_C$ [min$^{-1}$] | Competitor:miRNA complex decay rate | 1 | | | |
| $\delta$ [min$^{-1}$] | miRNA decay rate | 0.1 | | | |
| $\alpha_T$ [molecule min$^{-1}$] | $p_T$ synthesis rate | 0.5 | | | |
| $\delta_T$ [min$^{-1}$] | $p_T$ decay rate | 0.1 | | | |
| $\alpha_C$ [molecule min$^{-1}$] | $p_C$ synthesis rate | 0.5 | | | |
| $\delta_C$ [min$^{-1}$] | $p_C$ decay rate | 0.1 | | | |
| Parameter | Description | Fig. 2B,C | Figs 2E,F, 3 and 5A | Figs 5B,C and 6 | Fig. 4 |
| $\kappa_T$ [min$^{-1}$] | Target:miRNA complex catalytic decay rate | 0.001 | 0.001 | 0.001 | 0.001 (solid curves), 7.0 (dashed curves) |
| $k_C^+$ [molecule$^{-1}$ min$^{-1}$] | Competitor:miRNA complex association rate | 0 | see captions | see captions | see caption |
| $k^+$ [molecule$^{-1}$ min$^{-1}$] | Protein complex synthesis rate | 0 | 0 | 0.002 | 0 |
| $k^-$ [min$^{-1}$] | Protein complex dissociation rate | 0 | 0 | 0.001 | 0 |
| $\delta_P$ [min$^{-1}$] | Protein complex decay rate | 0 | 0 | 0.1 | 0 |

**Table 1. Parameters values used to obtain the figures.** Top table: parameters that do not change across the figures. Bottom table: remaining parameters with their figure-specific values. 'See caption(s)' indicates that the values of the corresponding parameters are reported in the figure captions. Parameters are chosen so as to keep the sizes of molecular populations in the range [0–1000].

$$\begin{aligned}
\frac{dm_i}{dt} &= -d_i m_i + b_i - k_i^+ \mu m_i + k_i^- c_i + \xi_i - \xi_i^+ + \xi_i^-, \\
\frac{dc_i}{dt} &= k_i^+ \mu m_i - (k_i^- + \kappa_i + \sigma_i) c_i + \xi_i^\sigma + \xi_i^+ - \xi_i^- - \xi_i^\kappa, \\
\frac{d\mu}{dt} &= -\delta\mu + \beta - \sum_i k_i^+ \mu m_i + \sum_i (k_i^- + \kappa_i) c_i + \xi_\mu - \sum_i \xi_i^+ + \sum_i \xi_i^- + \sum_i \xi_i^\kappa, \\
\frac{dp_i}{dt} &= \alpha_i m_i - \delta_i p_i - k^+ p_T p_C + k^- C_p + \xi_{p_i} + \xi^- - \xi^+, \\
\frac{dC_p}{dt} &= k^+ p_T p_C - k^- C_p - \delta_p C_p + \xi_{C_p} - \xi^- + \xi^+.
\end{aligned} \quad (3)$$

where $i \in \{C, T\}$. The different $\xi$-terms stand for the contributions to the noise coming from different processes. Each of these terms is assumed to have zero mean and variance given by

$$\begin{aligned}
\langle \xi_i(t)\xi_i(t')\rangle &= (d_i \bar{m}_i + b_i)\delta(t - t'), \\
\langle \xi_i^+(t)\xi_i^+(t')\rangle &= k_i^+ \bar{m}_i \bar{\mu}\delta(t - t'), \\
\langle \xi_i^-(t)\xi_i^-(t')\rangle &= k_i^- \bar{c}_i \delta(t - t'), \\
\langle \xi_\mu(t)\xi_\mu(t')\rangle &= (\delta\bar{\mu} + \beta)\delta(t - t'), \\
\langle \xi_i^\kappa(t)\xi_i^\kappa(t')\rangle &= \kappa_i \bar{c}_i \delta(t - t'), \\
\langle \xi_i^\sigma(t)\xi_i^\sigma(t')\rangle &= \sigma_i \bar{c}_i \delta(t - t'), \\
\langle \xi_{p_i}(t)\xi_{p_i}(t')\rangle &= (\alpha_i \bar{m}_i + \delta_i \bar{p}_i)\delta(t - t'), \\
\langle \xi^+(t)\xi^+(t')\rangle &= k^+ \bar{p}_T \bar{p}_C \delta(t - t'), \\
\langle \xi^-(t)\xi^-(t')\rangle &= k^- \bar{C}_p \delta(t - t'), \\
\langle \xi_{C_p}(t)\xi_{C_p}(t')\rangle &= \delta_p \bar{C}_p \delta(t - t'),
\end{aligned} \quad (4)$$

where the over-line stands for an average over time in the steady state. From the stationarity conditions, one finds in particular

$$\begin{align}
\overline{m}_i &= \frac{b_i + k_i^- \overline{c}_i}{d_i + k_i^+ \overline{\mu}}, \\
\overline{\mu} &= \frac{\beta + \sum_i (k_i^- + \kappa_i) \overline{c}_i}{\delta + \sum_i k_i^+ \overline{m}_i}, \\
\overline{c}_i &= \frac{k_i^+ \overline{\mu} \overline{m}_i}{\sigma_i + k_i^- + \kappa_i}, \\
\overline{p}_i &= \frac{\alpha_i \overline{m}_i + k^- \overline{C}_p}{\delta_i + k^+ \overline{p}_j} \qquad (i \neq j), \\
\overline{C}_p &= \frac{k^+ \overline{p_T p_C}}{k^- + \delta_p}.
\end{align}$$
(5)

By selectively setting some of the parameters appearing above to zero one easily obtains the equations describing the different circuits studied here.

**Linear Noise Approximation.** Let us denote the vector of molecular levels for the different species involved by $\mathbf{x} = \{m_T, m_C, \mu, c_T, c_C, p_T, p_T\}$ and by $\overline{\mathbf{x}}$ its steady state value. Assuming that the divergence from the steady state, $\delta \mathbf{x} = \mathbf{x} - \overline{\mathbf{x}}$, is small and expanding Eq. (3) around $\overline{\mathbf{x}}$, at the leading order one gets

$$\frac{d}{dt} \delta \mathbf{x} = \mathbf{A} \delta \mathbf{x} + \mathbf{x},$$
(6)

where $\mathbf{A}$ is the matrix of the first order derivatives evaluated at the steady state, while $\xi$ represents a white noise with zero mean and covariance matrix $\langle \xi_a(t) \xi_b(t') \rangle = \Gamma_{ab} \delta(t - t')$, where the indices $a$ and $b$ range over the components of $\mathbf{x}$ (non-zero elements of the covariance matrix are given by Eq. (4)). From Eq. (6) it follows that[43]

$$\langle \delta x_a \delta x_b \rangle = - \sum_{i,l,s,r} B_{as} B_{br} \frac{\Gamma_{il}}{\lambda_s + \lambda_r} B_{si}^{-1} B_{rl}^{-1},$$
(7)

where $B$'s and $\lambda$'s denote the eigenvectors and eigenvalues of $\mathbf{A}$. The above formula can be used to estimate correlations and Pearson coefficients of all molecular species involved in the system.

**Stochastic simulations.** We have employed the Gillespie algorithm (GA), a classical stochastic simulation method that computes the population dynamics of $N$ well-mixed molecular species interacting through one of $M$ reactions[42]. The dynamics of a biochemical system is obtained based on the probability $P(R, \tau)$ of an event of reaction $R$ to take place in the next time interval of size $\tau$. The latter can be calculated for every set of molecular populations given the chemical reaction rates[42]. In brief, the GA works as follows:

**Step 1** (**initialization**): Set up initial populations for all molecular species,
**Step 2:** Draw a pair $(R, \tau)$ from $P(R, \tau)$,
**Step 3:** Update molecular populations according to the selected reaction $R$ and advance time by $\tau$,
**Step 4:** Repeat Steps 2 and 3 until a pre-determined termination time is reached.

See Gibson *et al.*[74] for a more detailed presentation of the method.

A C++ implementation of the algorithm for the network shown in Fig. 7 is available at https://github.com/araksm/Protein-Expression-Noise.

**Protocol for computing capacities.** Capacities of the different regulatory channels we consider were computed as follows

**Step 1:** Calculate the output noise $\sigma_O(b_T)$ (variance of the output $O$) obtained at stationarity by the GA for any given value of the input variable $b_T$ in the range $(b_T^{\min}, b_T^{\max})$

**Step 2:** Compute the optimal input distribution

$$P_{\text{opt}}(x) = \frac{1}{Z} \sqrt{\frac{1}{\sigma_O^2(b_T)} \left( \frac{\partial \overline{O}(b_T)}{\partial b_T} \right)^2},$$
(8)

with $Z = \int_{b_T^{\min}}^{b_T^{\max}} db_T \sqrt{\frac{1}{\sigma_O^2(b_T)} \left( \frac{\partial \overline{O}(b_T)}{\partial b_T} \right)^2}$, where $\overline{O}(b_T)$ is the mean expression level of the output for the input $b_T$. Following[49,50], in the limit of small noise the channel's capacity coincides with the mutual information obtained when the input variable $b_T$ is distributed according to $P_{\text{opt}}(b_T)$.

**Step 3:** Generate an input signal according to $P_{\text{opt}}(b_T)$, record corresponding output signal $O(b_T)$ and calculate the joint probability distribution $P(b_T, O)$.

**Step 4:** Calculate the capacity $I_{\max}(b_T, O)$ from the data obtained in Step 3 via

$$I_{\max}(b_T, O) = \int_{b_T^{\min}}^{b_T^{\max}} db_T \int_{O^{\min}}^{O^{\max}} dO\, P(b_T, O) \log_2 \frac{P(b_T, O)}{P_{\text{opt}}(b_T) P(O)}, \tag{9}$$

where $P(O) = \int_{b_T^{\min}}^{b_T^{\max}} db_T\, P(b_T, O)$, while $O^{\min}$ and $O^{\max}$ denote maximum and minimum output expressed levels correspondingly.

## References

1. Shahrezaei, V. & Swain, P. S. Analytical distributions for stochastic gene expression. *PNAS* **105,** 17256–17261 (2008).
2. El-Samad, H. & Khammash, M. Regulated degradation is a mechanism for suppressing stochastic fluctuations in gene regulatory networks. *Biophys. J.* **90,** 3749–3761 (2006).
3. Singh, A. & Hespanha, J. P. Evolution of autoregulation in the presence of noise. *IET Syst. Biol.* **3,** 368–378 (2009).
4. Lestas, I., Vinnicombegv, G. & Paulsson, J. Fundamental limits on the suppression of molecular fluctuations. *Nature* **467,** 174–178 (2012).
5. Bundschuh, R., Hayot, F. & Jayaprakash, C. The role of dimerization in noise reduction of simple genetic networks. *J. Theor. Biol.* **220,** 261–269 (2003).
6. Pedraza, J. M. & Paulsson, J. Effects of molecular memory and bursting on fluctuations in gene expression. *Science* **319,** 339–343 (2008).
7. Morishita, Y. & Aihara, K. Noise-reduction through interaction in gene expression and biochemical reaction processes. *J. Theor. Biol.* **228,** 315–325 (2004).
8. Swain, P. S. Efficient attenuation of stochasticity in gene expression through post-transcriptional control. *J. Mol. Biol.* **344,** 956–976 (2004).
9. Bosia, C., Osella, M., Baroudi, M. E., Corà, D. & Caselle, M. Gene autoregulation via intronic microRNAs and its functions. *BMC Syst. Biol.* **6,** 131 (2012).
10. Mittal, N., Roy, N., Babu, M. M. & Jangaa, S. C. Dissecting the expression dynamics of RNA-binding proteins in posttranscriptional regulatory networks. *PNAS* **106,** 20300–20305 (2009).
11. Joshi, A., Beck, Y. & Michoel, T. Post-transcriptional regulatory networks play a key role in noise reduction that is conserved from micro-organisms to mammals. *FEBS J.* **279,** 3501–3512 (2012).
12. Bartel, D. P. MicroRNAs: genomics, biogenesis, mechanism, and function. *Cell* **116,** 281–297 (2004).
13. Chi, S. W., Zang, J. B., Mele, A. & Darnell, R. B. Argonaute HITS-CLIP decodes microRNA-mRNA interaction maps. *Nature* **460,** 479–486 (2009).
14. Poliseno, L. *et al.* A coding-independent function of gene and pseudogene mRNAs regulates tumour biology. *Nature* **465,** 1033–1038 (2010).
15. Cesana, M. *et al.* A long noncoding RNA controls muscle differentiation by functioning as a competing endogenous RNA. *Cell* **147,** 358–369 (2011).
16. Denzler, R., Agarwal, V., Stefano, J., Bartel, D. P. & Stoffel, M. Assessing the ceRNA Hypothesis with Quantitative Measurements of miRNA and Target Abundance. *Mol. Cell* **54,** 766–776 (2014).
17. Bosson, A. D., Zamudio, J. R. & Sharp, P. A. Endogenous miRNA and target concentrations determine susceptibility to potential ceRNA competition. *Mol. Cell* **56,** 347–359 (2015).
18. Ebert, M. S., Neilson, J. R. & Sharp, P. A. MicroRNA sponges: Competitive inhibitors of small RNAs in mammalian cells. *Nat. Methods* **4,** 721–726 (2007).
19. Das, J., Chakraborty, S., Podder, S. & Ghosh, T. C. Complex-forming proteins escape the robust regulations of miRNA in human. *FEBS Letters* **587,** 2284–2287 (2013).
20. Baek, D. *et al.* The impact of microRNAs on protein output. *Nature* **455,** 64–71 (2008).
21. Selbach, M. *et al.* Widespread changes in protein synthesis induced by microRNAs. *Nature* **455,** 58–63 (2008).
22. Riba, A., Bosia, C., El Baroudi, M., Ollino, L. & Caselle, M. A Combination of Transcriptional and MicroRNA Regulation Improves the Stability of the Relative Concentrations of Target Genes. *PloS Comput. Biol.* **10,** e1003490 (2014).
23. Gurtan, A. M. & Sharp, P. A. The Role of miRNAs in Regulating Gene Expression Networks. *J. Mol. Biol.* **425,** 3582–3600 (2013).
24. Schmiedel, J. M. *et al.* MicroRNA control of protein expression noise. *Science* **348,** 128–132 (2015).
25. Siciliano, V. *et al.* miRNAs confer phenotypic robustness to gene networks by suppressing biological noise. *Nat. Comm.* **4,** 2364 (2013).
26. Osella, M., Bosia, C., Corá, D. & Caselle, M. The role of incoherent microRNA-mediated feedforward loops in noise buffering. *PloS Comput. Biol.* **7,** e1001101 (2011).
27. Bosia, C., Pagnani, A. & Zecchina, R. Modelling Competing Endogenous RNA Networks. *PLoS ONE* **8,** e66609 (2013).
28. Figliuzzi, M., Marinari, E. & De Martino, A. MicroRNAs as a selective channel of communication between competing RNAs: a steady-state theory. *Biophys. J.* **104,** 1203–1213 (2013).
29. Salmena, L., Poliseno, L., Tay, Y., Kats, L. & Pandolfi, P. P. A ceRNA hypothesis: the Rosetta Stone of a hidden, RNA language? *Cell* **146,** 353–358 (2011).
30. Arvey, A., Larsson, E., Sander, C., Leslie, C. S. & Marks, D. S. Target mRNA abundance dilutes microRNA and siRNA activity. *Mol. Syst. Biol.* **6,** 363 (2010).
31. Franco-Zorrilla, J. M. *et al.* Target mimicry provides a new mechanism for regulation of microRNA activity. *Nature Genet.* **39,** 1033–1037 (2007).
32. Levine, E., Zhang, Z., Kuhlman, T. & Hwa, T. Quantitative Characteristics of Gene Regulation by Small RNA. *PLoS Biol.* **5,** e229 (2007).
33. Jens, M. & Rajewsky, N. Competition between target sites of regulators shapes post-transcriptional gene regulation. *Nat. Rev. Genet.* **16,** 113–126 (2015).
34. Ala, U. *et al.* Integrated transcriptional and competitive endogenous RNA networks are cross-regulated in permissive molecular environments. *PNAS* **110,** 7154–7159 (2013).
35. Karreth, F. A. *et al. In vivo* identification of tumor-suppressive PTEN ceRNAs in an oncogenic BRAF-induced mouse model of melanoma. *Cell* **147,** 382–395 (2011).
36. Martirosyan, A., Figliuzzi, M., Marinari, E. & De Martino, A. Probing the Limits to MicroRNA-Mediated Control of Gene Expression. *PLoS Comput. Biol.* **12,** e1004715 (2016).
37. Yuan, X. *et al.* Clustered microRNAs' coordination in regulating protein-protein interaction network. *BMC Syst. Biol.* **3,** 65 (2009).
38. Sass, S. *et al.* MicroRNAs coordinately regulate protein complexes. *BMC Syst. Biol.* **5,** 136 (2011).
39. Hsu, C. W., Juan, H.-F. & Huang, H.-C. Characterization of microRNA-regulated protein-protein interaction network. *Proteomics* **8,** 1975–1979 (2008).
40. Liang, H. & Li, W.-H. MicroRNA regulation of human protein-protein interaction network. *RNA* **13,** 1402–1408 (2007).
41. Zhu W. & Chen Y.-P. P. Computational developments in microRNA-regulated protein-protein interactions. *BMC Syst. Biol.* **8,** 14 (2014).

42. Gillespie, D. T. Exact stochastic simulation of coupled chemical reactions. *J. Phys. Chem. A* **81,** 2340–2361 (1977).
43. Swain, P. S. Efficient Attenuation of Stochasticity in Gene Expression Through Post-transcriptional Control. *J. Mol. Biol.* **344,** 965–976 (2004).
44. Mukherji, S. *et al.* MicroRNAs can generate thresholds in target gene expression. *Nat. Genet.* **43,** 854–859 (2011).
45. Levine, E. & Hwa, T. Small RNAs establish gene expression thresholds. *Curr. Opin. Microbiol.* **11,** 574–579 (2008).
46. Lai, X., Wolkenhauer, O. & Vera, J. Understanding microRNA-mediated gene regulatory networks through mathematical modelling. *Nucl. Acids. Res.* **44,** 6019–6035 (2016).
47. Figliuzzi, M., De Martino, A. & Marinari, E. RNA-based regulation: dynamics and response to perturbations of competing RNAs. *Biophys. J.* **107,** 1011–1022 (2014).
48. Tkačik, G., Walczak, A. M. & Bialek, W. Optimizing information flow in small genetic networks. *Phys. Rev. E* **80,** 031920 (2009).
49. Tkačik, G., Callan, C. G. Jr. & Bialek, W. Information Capacity of genetic regulatory networks. *Phys. Rev. E* **78,** 011910 (2008).
50. Tkačik, G. & Walczak, A. M. Information transmission in genetic regulatory networks: a review. *J. Phys. Condens. Matter* **23,** 153102 (2011).
51. Chekulaeva, M. & Filipowicz, W. Mechanisms of miRNA-mediated post-transcriptional regulation in animal cells. *Curr. Opin. Chem. Biol.* **21,** 452–460 (2009).
52. Valencia-Sanchez, M. A., Liu, J., Hannon, G. J. & Parker, R. Control of translation and mRNA degradation by miRNAs and siRNAs. *Genes. Dev.* **20,** 515–524 (2006).
53. Muers, M. Small RNAs: Recycling for silencing. *Nat. Rev. Genet.* **12,** 227 (2011).
54. Hutvágner, G. & Zamore, P. D. A microRNA in a multiple-turnover RNAi enzyme complex. *Science* **297,** 2056–2060 (2002).
55. Zeng, Y., Wagner, E. J. & Cullen, B. R. Both natural and designed micro RNAs can inhibit the expression of cognate mRNAs when expressed in human cells. *Mol. Cell* **9,** 1327–1333 (2002).
56. Zeng, Y., Yi, R. & Cullen, B. R. MicroRNAs and small interfering RNAs can inhibit mRNA expression by similar mechanisms. *PNAS* **100,** 9779–9784 (2003).
57. Doench, J. G., Petersen, C. P. & Sharp, P. A. siRNAs can function as miRNAs. *Genes Dev.* **17,** 438–442 (2003).
58. Du, B. *et al.* MicroRNA-545 Suppresses Cell Proliferation by Targeting Cyclin D1 and CDK4 in Lung Cancer Cells. *PLoS One* **9,** e88022 (2014).
59. Nadal, A. *et al.* Association of CDK4 and CCND1 mRNA overexpression in laryngeal squamous cell carcinomas occurs without CDK4 amplification. *Virchows Arch.* **450,** 161–167 (2007).
60. Kwon, J. *et al.* Integrin alpha 6: A novel therapeutic target in esophageal squamous cell carcinoma. *Int. J. Oncol.* **43,** 1523–1530 (2013).
61. Bartel, D. P. & Chen, C. Z. Micromanagers of gene expression: the potentially widespread influence of metazoan microRNAs. *Nat. Rev. Genet.* **5,** 396–400 (2004).
62. Shi, Y. *et al.* MicroRNA Regulation of Neural Stem Cells and Neurogenesis. *J. Neurosci.* **30,** 14931–14936 (2010).
63. Klein, M. E., Impey, S. & Goodman, R. H. Role reversal: the regulation of neuronal gene expression by microRNAs. *Curr. Opin. Neurobiol.* **15,** 507–513 (2005).
64. Giraldez, A. J. *et al.* MicroRNAs regulate brain morphogenesis in zebrafish. *Science* **308,** 833–838 (2005).
65. Naguibneva, I. *et al.* The microRNA miR-181 targets the homeobox protein Hox-A11 during mammalian myoblast differentiation. *Nat. Cell Biol.* **8,** 278–284 (2006).
66. Hatfield, S. D. *et al.* Stem cell division is regulated by the microRNA pathway. *Nature* **435,** 974–978 (2005).
67. Lynn, F. C. Meta-regulation: microRNA regulation of glucose and lipid metabolism. *Trends Endocrinol. Metab* **20,** 452–459 (2009).
68. Skalsky, R. L. & Cullen, B. R. Viruses, microRNAs, and host interactions. *Annu. Rev. Microbiol.* **64,** 123–141 (2010).
69. Breda, J., Rzepiela, A. J., Gumienny, R., van Nimwegen, E. & Zavolan, M. Quantifying the strength of miRNA-target interactions. *Methods* **85,** 90–99 (2015).
70. Bosia, C. *et al.* Quantitative study of crossregulation, noise and synchronization between microRNA targets in single cells; Preprint. Available: https://arxiv.org/abs/1503.06696 arXiv:1503.06696 (2015).
71. Alshalalfa, M., Bader, G. D., Bismar, T. A. & Alhajj, R. Coordinate MicroRNA-Mediated Regulation of Protein Complexes in Prostate Cancer. *PLoS One* **8,** e84261 (2013).
72. Park, S.-M., Gaur, A. B., Lengyel, E. & Peter, M. E. The miR-200 family determines the epithelial phenotype of cancer cells by targeting the E-cadherin repressors. *Genes Dev.* **22,** 894–907 (2008).
73. Huelsken, J. & Behrens, J. The Wnt signalling pathway. *J. Cell Sci.* **115,** 3977–3978 (2002).
74. Gibson, M. A. & Bruck, J. Efficient Exact Stochastic Simulation of Chemical Systems with Many Species and Many Channels. *J. Phys. Chem. A* **104,** 1876–1889 (2000).

### Acknowledgements
We acknowledge support from the Marie Curie Action ITN NETADIS (FP7/Grant 290038, http://netadis.eu), the joint IIT–Sapienza Lab "Nanomedicine" (http://lns.iit.it/), MIUR FIRB 2008 program (http://futuroinricerca.miur.it/) and MIUR PRIN project "Statistical mechanics of disordered complex systems" (http://www.istruzione.it/).

### Author Contributions
A.D.M., A.P. and E.M. conceived and designed the experiments. A.M. performed the experiments. All authors analyzed the data and wrote the manuscript.

### Additional Information
**Competing Interests:** The authors declare no competing financial interests.

**How to cite this article:** Martirosyan, A. *et al.* ceRNA crosstalk stabilizes protein expression and affects the correlation pattern of interacting proteins. *Sci. Rep.* **7,** 43673; doi: 10.1038/srep43673 (2017).

**Publisher's note:** Springer Nature remains neutral with regard to jurisdictional claims in published maps and institutional affiliations.